\documentstyle[11pt]{article}
\begin{document}
%---------------------------------------------------
\title{An Electroweak Model Without Higgs particle}
\author{{Ning Wu}\\
{\small CCAST (World Lab), P.O.Box 8730, Beijing 100080, P.R.China}\\
{\small and}\\ 
{\small Division 1, Institute of High Energy Physics, P.O.Box 918-1, 
Beijing 100039, P.R.China}}
\maketitle
\vskip 0.8in

\noindent
PACS Numbers: 12.15-y, 11.15-q,  11.10-z,  12.10-g\\
Keywords: electroweak interactions,  gauge symmetry,  gauge field, 
symmetry breaking,  long-range force  \\
\vskip 0.8in

\noindent
\begin{abstract} 
In this paper, we try to construct an electroweak model which 
aviods using Higgs mechanism. We 
simultaneously introduce two sets of gauge bosons so as to keep 
the masses of gauge bosons 
$W^{\pm}$ and $Z$ non-zero without Higgs mechanism. 
In order to introduce symmetry breaking,
we need a vacuum potential.
In a proper limit, 
this model will approximately
return to the standard model, except that there exists no Higgs particle 
and couplings between 
Higgs particle and fermions or gauge fields. The fundamental dynamical characteristics of this model 
are similar to those of the standard model. 
\end{abstract}

\newpage

%-------------------------------------------------------
Since Yang and Mills founded non-Abel gauge field theory in 1954 
\lbrack 1 \rbrack, gauge field theory has been extensively applied to 
elementary particle theory.
 But we know that, if gauge symmetry is strictly preserved in 
the Yang-Mills gauge field theory, the masses of gauge bosons should be zero. The introduction
of spontaneously symmetry breaking and  Higgs mechanism make it possible for 
physicists to construct electroweak model by using gauge theory. In the sixties, Glashow \lbrack 2 
\rbrack , Weinberg \lbrack 3 \rbrack and Salam \lbrack 4 \rbrack 
founded the unified electroweak standard 
model. The gauge bosons $W^{\pm}$ and $Z$ which are predicted by the standard model have 
been found by experiment. 
But Higgs particle, which is necessary for the standard model, has not been
found until now \lbrack 5 \rbrack .\\

Using the gauge field theory which has massive gauge bosons\lbrack 6-7
\rbrack, we could construct an 
electroweak model without using Higgs mechanism. The fundamental 
characteristics of the new electroweak model are: (1) there exists no 
Higgs particle in the theory; (2) 
the interaction properties of the new theory are the same as those 
of the standard model, and in a 
proper limit, this model will approximately return to the standard 
model; and (3) there exist two sets of gauge bosons 
in the theory, one set contains $W^{\pm} ,~ Z$ and photon which we are familiar with, another set 
contains only massless gauge bosons. In 
the new electroweak model, some physical arguments, such as the structures of neutral current and 
charged currents, the masses of intermediate bosons $W^{\pm}$ and $Z$ , the coupling constant of 
electromagnetic interactions , etc., are completely the same as those in the standard model. In a 
word, except for Higgs particle and those terms concern Higgs particle, all other properties are 
inherited by the new model.   \\

 Now, let's discuss the electroweak interactions of leptons. Let $e$ represent $e,\mu$ or $\tau$, 
and $\nu$ represent the corresponding neutrinos $\nu _e,\nu_{\mu}$ or $\nu _{\tau}$. Suppose that 
leptons $e$ and $\nu$ form a left-hand doublet $\psi_L$ and a right-hand singlet $e_R$. The 
definitions of $\psi_L$ and $e_R$ are the same as those in standard model. At the same time, in 
order 
to obtain the masses of $W^{\pm}$ and $Z$, we need four kinds of gauge fields. There are two kinds 
of gauge fields $F_{1 \mu}$ and $F_{2 \mu}$ which correspond to $SU(2)_L$ symmetry and two 
kinds of gauge fields $B_{1 \mu}$ and $B_{2 \mu}$ which correspond to $U(1)_Y$ symmetry. \\

 Usually in lagrangian, the mass terms have the following forms
$$
-\frac{\mu ^2}{2} \phi (x) \phi(x) ~~,~~ -m \overline{\psi}(x) \psi (x)
\eqno{(1)} 
$$
where $\mu$ and $m$ are mass parameters. If we hope that there exist no parameters which have 
mass dimension, we could change the above terms into the following forms
$$
-\frac{1}{2} v^{\dag } \phi (x) \phi(x) v ~~,~~ - \overline{\psi}(x)  v  \psi (x)
\eqno{(2)} 
$$
with $v$ a potential which has mass dimension. We name it a vacuum potential for the moment. The 
vacuum potential $v$ has no dynamical terms. 
We know that the $SU(2)_L \times U(1)_Y$ symmetry is an approximate symmetry. Later, we will 
introduce symmetry breaking through vacuum potential $v$.\\

The lagrangian of the model is 
$$
{\cal L} = {\cal L} _l + {\cal L} _g + {\cal L} _{v-l}
\eqno{(3)} 
$$
$$
{\cal L }_l= - \overline{\psi}_L \gamma ^{\mu} 
(\partial _{\mu}+ \frac{i}{2}g \prime B_{1 \mu} -ig F_{1 \mu} ) \psi _L
- \overline{e}_R \gamma ^{\mu} 
(\partial _{\mu}+ ig \prime B_{1 \mu} ) e_R 
\eqno{(4)} 
$$
$$
\begin{array}{ccl}
{\cal L}_g &= &-\frac{1}{4}  F^{i \mu \nu}_1 F^i_{1 \mu \nu} 
- \frac{1}{4} F^{i \mu \nu}_2 F^i_{2 \mu \nu} 
-\frac{1}{4}  B^{\mu \nu}_1 B_{1 \mu \nu}  
-\frac{1}{4}  B^{\mu \nu}_2 B_{2 \mu \nu}  \\
&& - v^{\dag} 
\left \lbrack  {\rm cos} \theta _W ( {\rm cos} \alpha F_1^{\mu}+{\rm sin}\alpha F_2^{\mu}) -
{\rm sin}\theta _W ( {\rm cos}\alpha B_1^{\mu}+{\rm sin}\alpha B_2^{\mu} ) \right \rbrack \\
&&~\cdot 
\left \lbrack  {\rm cos} \theta _W ( {\rm cos} \alpha F_{1 \mu}+{\rm sin}\alpha F_{2 \mu}) -{\rm 
sin}\theta _W ( {\rm cos}\alpha B_{1 \mu}+{\rm sin}\alpha B_{2 \mu} ) \right \rbrack
v
\end{array}
\eqno{(5)} 
$$
$$
{\cal L} _{v-l} = -f (\overline{e}_R v^{\dag} \psi _L +\overline{\psi}_L v e_R) 
\eqno{(6)} 
$$
where
$$
F_{1 \mu \nu}^i = \partial _{\mu} F_{1 \nu}^i - \partial _{\nu} F_{1 \mu}^i
+g \epsilon _{ijk} F_{1 \mu}^j    F_{1 \nu}^k
\eqno{(7a)} 
$$
$$
F_{2 \mu \nu}^i = \partial _{\mu} F_{2 \nu}^i - \partial _{\nu} F_{2 \mu}^i
-g {\rm tg} \alpha \epsilon _{ijk} F_{2 \mu}^j    F_{2 \nu}^k
\eqno{(7b)} 
$$
$$
B_{m \mu \nu}= \partial _{\mu} B_{m \nu}- \partial _{\nu} B_{m \mu}
~~~ (m=1,2). 
\eqno{(8)} 
$$
The lagrangian density defined by eqs.(3-6) has strict local $SU(2)_L \times U(1)_Y$ gauge 
symmetry.\\

The symmetry breaking of the model is accomplished through the change of the vacuum 
potential. We could suppose that the lagrangian defined by eq.(3-6) describes a state of matter which 
exists in a 
condition of extreme high temperature. This state is a special phase of vacuum and may exist at a 
moment of Big Bang. Alone with the decreasing of the temperature of the state, the phase transition 
of the vacuum occurs. After phase transition, the vacuum potential changes into:
$$
v =\left ( 
\begin{array}{c}
0\\
\mu / \sqrt{2}
\end{array}
\right ).
\eqno{(9)} 
$$

 Gauge fields $F_{1 \mu}, F_{2 \mu}, B_{1 \mu}$ and $B_{2 \mu}$ are not eigenvectors of 
mass matrix. In order to obtain the eigenvectors of mass matrix, let's first make the following 
transformations:
$$
W_{\mu}={\rm cos}\alpha F_{1 \mu}+{\rm sin}\alpha F_{2 \mu}
\eqno{(10)} 
$$
$$
W_{2 \mu}=-{\rm sin}\alpha F_{1 \mu}+{\rm cos}\alpha F_{2 \mu}
\eqno{(11)} 
$$
$$
C_{1 \mu}={\rm cos}\alpha B_{1 \mu}+{\rm sin}\alpha B_{2 \mu}
\eqno{(12)} 
$$
$$
C_{2 \mu}=-{\rm sin}\alpha B_{1 \mu}+{\rm cos}\alpha B_{2 \mu} ,
\eqno{(13)} 
$$
then make   transformations:
$$
Z_{\mu}= {\rm sin}\theta _W C_{1 \mu}-{\rm cos}\theta _W W^3_{ \mu}
\eqno{(14)} 
$$
$$
A_{\mu}= {\rm cos}\theta _W C_{1 \mu}+{\rm sin}\theta _W W^3_{ \mu}
\eqno{(15)} 
$$
$$
Z_{2 \mu}= {\rm sin}\theta _W C_{2 \mu}-{\rm cos}\theta _W W^3_{2 \mu}
\eqno{(16)} 
$$
$$
A_{2 \mu}= {\rm cos}\theta _W C_{2 \mu}+{\rm sin}\theta _W W^3_{2 \mu}. 
\eqno{(17)} 
$$
After the above transformations, the lagrangian densities change into:
$$
\begin{array}{ccl}
{\cal L}_l +{\cal L}_{v-l} &= & - \overline{e} (\gamma ^{\mu} 
\partial _{\mu}+ \frac{1}{\sqrt{2}} f \mu ) e 
-\overline{\nu}_L \gamma ^{\mu} \partial _{\mu}\nu _L\\
&&
+\frac{1}{2} \sqrt{g^2 + {g \prime}^2} {\rm sin}2\theta_W  
j^{em}_{\mu} 
 ( {\rm cos}\alpha A^{\mu}- {\rm sin}\alpha  A^{\mu}_2  )  \\
&&
- \sqrt{g^2 + {g \prime}^2} j^{z}_{\mu} 
( {\rm cos}\alpha Z^{\mu} - {\rm sin}\alpha Z_2^{ \mu} ) \\
&&
+ \frac{\sqrt{2}}{2} ig \overline{\nu}_L \gamma ^{\mu} e_L 
( {\rm cos}\alpha W_{\mu}^{+} - {\rm sin}\alpha W_{2 \mu}^{+} )  \\
&&
+ \frac{\sqrt{2}}{2} ig \overline{e}_L \gamma ^{\mu} {\nu}_L  
( {\rm cos}\alpha W_{\mu}^{-} - {\rm sin}\alpha W_{2 \mu}^{-} )
\end{array} 
\eqno{(3.12)} 
$$
$$
\begin{array}{ccl}
{\cal L}_g &= &-\frac{1}{2}  W^{+ \mu \nu}_0  W^{-}_{0 \mu \nu} 
-\frac{1}{4}  Z^{\mu \nu} Z_{ \mu \nu} 
-\frac{1}{4}  A^{\mu \nu} A_{ \mu \nu} 
 \\
&&-\frac{1}{2}  W^{+ \mu \nu}_{2 0}  W^{-}_{2 0 \mu \nu} 
-\frac{1}{4}  Z^{\mu \nu}_2  Z_{2  \mu \nu} 
-\frac{1}{4}  A^{\mu \nu}_2  A_{2  \mu \nu} 
\\
&& -\frac{\mu ^2}{2}  Z^{\mu }  Z_{ \mu } 
-\mu ^2 {\rm cos}^2 \theta _W  W^{+ \mu}  W^{-}_{\mu} +{\cal L}_{g I}
\end{array} ,
\eqno{(19)} 
$$
where, ${\cal L}_{g I}$ only contains interaction terms of gauge fields. In the above 
relations, we have used the following notations: if $A_{\mu}$ is a non-Abel gauge field, then
$$
A_{0 \mu \nu} = \partial _{\mu} A_{\nu} - \partial _{\nu} A_{\mu} ;
\eqno{(20)} 
$$
if $A_{\mu}$ is an Abel gauge field, then
$$
A_{ \mu \nu} = \partial _{\mu} A_{\nu} - \partial _{\nu} A_{\mu} ;
\eqno{(21)} 
$$
and
$$
W^{\pm}_{m \mu } = \frac{1}{\sqrt{2}} (W^1_{m \mu} \mp i W^2_{m \mu} )
~~~( m=1,2, ~W_{1 \mu} \equiv W_{\mu}  ) ,
\eqno{(22)} 
$$
$$
j_{ \mu }^{em} = -i \overline{e} \gamma _{\mu} e ,
\eqno{(23)} 
$$
$$
j_{ \mu }^{Z} = j^3_{\mu} - {\rm sin}^2 \theta_W  j^{em}_{\mu}. 
\eqno{(24)} 
$$

From eqs (18-19), we know that, the mass of lepton $e$ is $\frac{1}{\sqrt{2}} f \mu$, the 
mass of neutrino is still zero, the masses of the charged gauge bosons $W^{\pm}$ are $m_W = \mu 
{\rm cos} \theta$, the mass of neutral gauge boson $Z$ is 
$m_2= \mu = \frac{m_W}{{\rm cos} \theta _W}$ and 
the masses of gauge fields $A, A_2, Z_2, W_2^{\pm}$ are all zero. The current structures in the 
model are completely the same as those in the standard model. At the same time, we notice
that, there exist two different electromagnetic fields $A_{\mu}$ and $ A_{2 \mu}$, so there exist 
two different coupling constants of electromagnetic interactions. 
$$
e_1 = \frac{g  g'}{\sqrt{g^2 + {g'} ^2}} {\rm cos}\alpha
~~,~~
e_2 = \frac{g  g'}{\sqrt{g^2 + {g' }^2}} {\rm sin}\alpha .
\eqno{(25)} 
$$
The real electromagnetic field in nature should be a mixture of these two electromagnetic fields 
$A_{\mu}$ and $A_{2 \mu}$, so the effective coupling 
constant of the electromagnetic interactions should be
$$
e^2 = e_1^2 +e_2^2
~~,~~
e= \frac{g  g'}{\sqrt{g^2 + {g' }^2}} .
\eqno{(26)} 
$$
From this relation, we see that the value of the parameter 
$\alpha$ doesn't  affect the value of the 
effective coulping constant of  electromagnetic interactions. 
Once again, we have noticed that the 
expression of effective coupling constant $e$ of the 
electromagnetic interactions is completely the 
same as that in the standard model. \\

 If the parameter $\alpha$ is small, then in the leading term
approximation, except for the terms concern 
Higgs particle, the lagrangian density defined by eqs(18-19) will return to the lagrangian density of 
the standard model. Because the standard model is consonant well with high
energy experiments, we could anticipate that the parameter 
$\alpha$ is small. \\

 Experimentally speaking, if these five kinds of massless gauge bosons introduced in this paper 
exist in nature, some of them  might be regarded as $\gamma$ photon. The gauge boson $Z_2$ 
have similar interaction properties to those of  $\gamma$ photon, and they have  the same mass, so 
it is hard to distinguish between $\gamma$ photon and $Z_2$ boson in
experiment. If experimental physicists find 
that $\gamma$ 
photon takes part in weak interactions, that means that there exists 
$Z_2$ boson mixed in $\gamma$ photon.
Those charged massless gauge bosons $W_2^{\pm}$ might be regarded 
as charged photon or other
light charged particles. (Such as electron suppose that simultaneously create an invisible neutrino) 
It is an important work to verify the existence of these massless gauge bosons. \\

 Just revised the standard model and introduce another set of gauge bosons,
we could found a theory which simultaneously contains Higgs particle 
and those massless
gauge bosons. So, Higgs particle and massless gauge bosons could exist in
the same theory. But, if these massless gauge bosons were
found by experiment, Higgs particle is no longer needed in thoery. The
reason for the introduction of Higgs particle does not exist. \\

 Some properties of vacuum potential are similar to those of 
Higgs field, but
they have essential differences. The most important difference is that vacuum
potential has no kinematical energy term in the lagrangian. 
It has no dynamics. All these
are properties of vacuum, so we name it vacuum potential. We could consider
it from another point of view. We know that vacuum can be regarded as sea of
virtual particle. It has interactions with physical particle. 
This kind of interactions change the mass of
physical particle(see for example mass renormalization). So, if a naked particle
is massless, it could obtain mass through its interactions with vacuum. That
is the physical interpretation of eq(2) and could be regarded particially as
the origin of mass. The properties of vacuum are not stable forever. With
the change of the properties of vacuum, the symmetry of physical system will
change accordingly and the symmetry breaking occurs \lbrack 9 \rbrack. In this 
paper, we use v to denote the influences of vacuum to physical fields. \\

 We have clearly known that there exist two kinds of long-range force fields in nature: 
electromagnetic field and gravitation field. According to this paper, there may exist the third  
long-range force field. The third long-range force field is weak field which is transmitted by 
$Z_2$ boson. The influences of  weak long-range force  may be weakened by the adsorption of 
neutrino in matter, because the adsorption of neutrino in matter will make 
the object 
in a state of weak charge neutral. This long-range force might have
 some influences in cosmology. 
\\

We could construct a lot of electroweak models by using Higgs mechanism, but we could
construct only one electroweak model by using vacuum potential. So, 
in the new theory, the 
interaction properties are more fixed by the requirement of gauge symmetry. This property is 
important theoretically. \\

\section*{Reference:}
\begin{description}
\item[\lbrack 1 \rbrack]  C.N.Yang, R.L.Mills, Phys Rev {\bf 96} (1954) 191
\item[\lbrack 2 \rbrack]  S.Glashow, Nucl Phys {\bf 22}(1961) 579
\item[\lbrack 3 \rbrack]  S.Weinberg, Phys Rev Lett {\bf 19} (1967) 1264
\item[\lbrack 4 \rbrack]  A.Salam, in Elementary Particle Theory, eds.N.Svartholm(Almquist and 
Forlag, Stockholm,1968)
\item[\lbrack 5 \rbrack]  Particle Data Group, phys. Rev. {\bf d54} (1996) 20
\item[\lbrack 6 \rbrack]  Ning Wu, Gauge Field Theory With Massive Gauge Bosons,
hep-ph/9802236
\item[\lbrack 7 \rbrack]  Ning Wu, General gauge field theory, 
hep-ph/9805453
\end{description}

\end{document}